\documentclass[letterpaper,twocolumn,10pt]{article}
\PassOptionsToPackage{hidelinks}{hyperref}
\usepackage{usenix}

\usepackage{tikz}
\usetikzlibrary{arrows.meta,positioning,fit,backgrounds}
\usepackage{amsmath}
\usepackage{amssymb}
\usepackage{graphicx}
\usepackage{booktabs}
\usepackage{multirow}
\usepackage{xcolor}
\usepackage{hyperref}
\usepackage{algorithm}
\usepackage{algpseudocode}
\usepackage{pifont}
\newcommand{\cmark}{\ding{51}}
\newcommand{\xmark}{\ding{55}}

\begin{document}

\date{}

\title{\Large \bf Embedding Asymmetry as Attack Signal: Detecting and Defending Active Label Inference in Vertical Federated Learning}



\author{
{\rm Soumya Swagata Biswas}\\
 Bangladesh University of\\Engineering and Technology\\
\and
{\rm Dr. Muhammad Abdullah Adnan}\\
Bangladesh University of\\Engineering and Technology
}


\maketitle

\begin{abstract}
Vertical federated learning lets organizations train a shared model over features they hold separately, while one party keeps the labels private. The active label inference attack of Fu et al.~\cite{fu2022label} breaks this: a feature-holding party runs a modified optimizer, MaliciousSGD, that forces class structure into its embeddings and then recovers the labels with semi-supervised learning. Defenses that perturb the shared gradient fail, because the attack acts on the party's internal optimizer, not on the message it sends.

We show the attack leaves a signal the server can read from the embeddings it already receives. The adversary's Fisher class-separability pulls away from the honest party's, and the gap is visible within a few epochs. A defense triggered by this signal acts on the adversary alone. Scaling its gradient down handles few-class tasks, and a norm non-expansive projection handles many-class tasks, with the Discriminative Coverage Ratio explaining why the projection rank must grow with the class count. The defense breaks the alignment between embedding clusters and labels rather than destroying the embedding. Across four seeds on CIFAR-10, CIFAR-100, and CINIC-10, label inference falls below what honest training already leaks, at under two points of task-accuracy cost.
\end{abstract}

\section{Introduction}

Organizations often want to train a model over data they are not allowed to pool. In vertical federated learning (VFL), several parties hold different features for the same set of users, and one party also holds the labels~\cite{mcmahan2017fedavg,yang2019federated,kairouz2021advances}. A bank and an online retailer may share many customers while recording different things about them. A hospital and an insurer may share patients. The parties train a joint model without moving raw data: each party maps its own features to an embedding, a server combines the embeddings and produces a prediction, and gradients flow back to each party. The labels stay with one party, and they are often the most sensitive part of the data, for example whether a patient has a given disease or whether a borrower has defaulted. Keeping the labels private from the other parties is a stated goal of the setting.

Fu et al.~\cite{fu2022label} showed that a curious feature party can recover these labels anyway. Their strongest method is an active attack. Instead of only following the protocol, the attacker replaces its local optimizer with MaliciousSGD, which amplifies the attacker's own parameter updates during training. Over many epochs this pushes the attacker's embeddings into tight, well-separated class clusters, even though the attacker never sees a label. After training, the attacker attaches a small classifier to its saved embedding model and trains it with semi-supervised learning, using only a few dozen labeled examples. On CIFAR-10 this recovers the labels of about 95\% of the training set.

The attack is hard to stop because it works through the attacker's own optimizer, not through the messages the attacker sends. The usual protections perturb the gradient that travels between the parties: differential privacy adds noise~\cite{dwork2006calibrating}, and gradient compression drops small entries~\cite{lin2018dgc}. Both act on the communicated tensor. MaliciousSGD amplifies the internal parameter gradient, which is a different tensor derived from the communicated one through the attacker's own network. Perturbing the message slows the attacker but does not remove the class structure that the internal amplification keeps building. These defenses are also symmetric. They perturb the honest party as much as the attacker, so they trade real task accuracy for partial protection.

We start from a different question. Before perturbing anything, can the server tell that an attack is under way? It can. MaliciousSGD leaves a fingerprint in the embedding space: the attacker's embeddings become more class-separable than the honest party's. We measure separability with the Fisher ratio, the between-class scatter divided by the within-class scatter, computed per party from the embeddings the server already receives. We then watch the difference between the two parties. In honest training the difference stays near zero. Under the active attack it rises and stays positive, and in our experiments it crosses a fixed threshold by the fourth epoch on CIFAR-10 and by the seventh on CIFAR-100. The server computes this from the forward embeddings alone, with no access to any party's weights or optimizer.

The signal tells the server both when to act and on whom. Because only the attacker's separability is abnormal, the server can respond on the attacker's gradient and leave the honest party's gradient unchanged, which keeps the cost to the shared task small. The right response depends on how many classes the task has. With few classes, scaling the attacker's gradient down while the signal is high keeps its embeddings from settling into usable clusters. With many classes this is not enough. There the server removes the discriminative direction from the attacker's gradient with a projection. The projection is non-expansive: it can only shrink the gradient, never grow it, so $\lVert Pg\rVert \le \lVert g\rVert$. This bound matters. An earlier design of ours pushed back against the discriminative direction instead of only removing it, and under MaliciousSGD's amplification the gradient diverged to not-a-number and destroyed the model. A projection cannot do that.

The split between the two settings is not arbitrary. Removing one direction covers a fixed slice of a discriminative subspace whose size grows with the number of classes. We make this precise with the Discriminative Coverage Ratio, $\mathrm{DCR}(K,C) = K / (C-1)$, where $K$ is the number of directions the projection removes and $C$ is the number of classes. For ten classes, one direction covers about a ninth of the subspace. For a hundred classes, it covers about a hundredth. The ratio predicts that the projection rank must grow with the class count to hold coverage steady, and it explains why one setting of the defense transfers across the ten-class datasets we test while a single-direction projection is weak on a hundred classes.

We also looked at how the defense protects the labels, and the answer is not the obvious one. The projection does not erase the geometry of the attacker's embedding space. The clusters are still present, and by some measures they are even tighter than under an undefended attack. What breaks is the match between those clusters and the true labels. Semi-supervised model completion works by spreading a few labels across the embedding geometry, so it fails when the geometry no longer tracks the labels. This is not the same as adding noise to destroy the embedding, and it is why the defense holds task accuracy while cutting leakage.

We evaluate on CIFAR-10, CIFAR-100, and CINIC-10, over four random seeds each. Label inference falls from 94.95\% to 81.80\% on CIFAR-10, from 49.87\% to 23.44\% on CIFAR-100, and from 86.51\% to 62.72\% on CINIC-10. In every case the defended leakage sits below what honest training already leaks, and the shared task loses under two points of accuracy. We also report the defenses that did not work, including gradient noise, embedding noise, sign flipping, and an unbounded direction-reversal method, because those failures shaped the final design.

This paper makes the following contributions.
\begin{itemize}
\item We identify a server-side detection signal for the active label inference attack. The Fisher separability of the attacker's embeddings diverges from the honest party's, the gap is visible within a few epochs, and the server needs no access to any party's internals to see it.
\item We build an asymmetric defense that the signal triggers. It acts on the attacker's gradient alone and comes in two bounded forms, adaptive suppression for few-class tasks and a non-expansive projection for many-class tasks, validated on three datasets over four seeds each, with defended leakage below honest-training leakage at under two points of task cost.
\item We introduce the Discriminative Coverage Ratio, which explains why the projection rank must scale with the number of classes and predicts when a single-direction projection will fall short.
\item We show that the defense works by breaking the alignment between embedding clusters and true labels rather than by destroying embedding geometry, and we report a set of gradient-space and embedding-space defenses that fail, with the reason each one fails.
\end{itemize}

\section{Background and Threat Model}

\subsection{Vertical federated learning}
\begin{figure}[t]
\centering
\begin{tikzpicture}[font=\small,
  box/.style={draw,rounded corners,align=center,inner sep=4pt},
  ar/.style={-{Stealth[length=2mm]},thick},
  ard/.style={-{Stealth[length=2mm]},thick,densely dashed}]
  \node[box,fill=gray!12,minimum width=6.4cm,minimum height=1.05cm] (srv) at (0,3.0)
    {\textbf{Server (label party)}: top model, labels $y$\\[1pt]
     detector $\Delta_F = J_A - J_B$ \; | \; response on $g_A$};
  \node[box,fill=gray!6,minimum width=3.0cm,minimum height=1.0cm] (A) at (-1.95,0.5)
    {Party A (adversary)\\ $f_A$ + MaliciousSGD};
  \node[box,fill=gray!6,minimum width=3.0cm,minimum height=1.0cm] (B) at (1.95,0.5)
    {Party B (honest)\\ $f_B$ + SGD};
  \node[box,fill=white,minimum width=1.3cm] (def) at (-1.55,1.75) {defense};
  \node (xa) at (-1.95,-0.85) {$x_A$};
  \node (xb) at (1.95,-0.85) {$x_B$};
  \draw[ar] (xa) -- (A);
  \draw[ar] (xb) -- (B);
  \draw[ar] ([xshift=-3mm]A.north) -- node[left,pos=0.55]{$z_A$} ([xshift=-3mm]A.north |- srv.south);
  \draw[ar] ([xshift=3mm]B.north)  -- node[right,pos=0.55]{$z_B$} ([xshift=3mm]B.north |- srv.south);
  \draw[ard] ([xshift=3mm]A.north |- srv.south) -- (def.north);
  \draw[ard] (def.south) -- node[right,pos=0.5]{$g_A$} ([xshift=3mm]A.north);
  \draw[ard] ([xshift=-3mm]B.north |- srv.south) -- node[left,pos=0.5]{$g_B$} ([xshift=-3mm]B.north);
\end{tikzpicture}
\caption{Two-party VFL with the server-side defense. Solid arrows are forward embeddings; dashed arrows are returned gradients. Party~A runs MaliciousSGD. The server measures the Fisher divergence $\Delta_F$ and, when it fires, modifies only $g_A$. Party~B's gradient is untouched.}
\label{fig:system}
\end{figure}
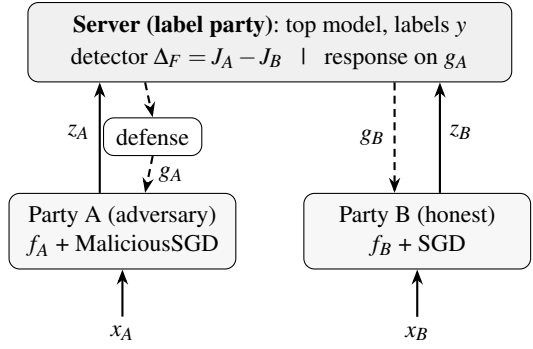

In vertical federated learning, a fixed set of users is shared across parties, but each party holds different features for those users. We study the two-party case used by prior work~\cite{fu2022label}. Party~A and Party~B each hold half of the input features. A server holds the labels and a top model. Each party runs a bottom model $f_A$ or $f_B$ that maps its own features $x_A$ or $x_B$ to an embedding $z_A = f_A(x_A)$ or $z_B = f_B(x_B)$. On each batch, both parties send their embeddings to the server. The server passes the concatenated embeddings through the top model, computes the loss against the labels it holds, and sends each party the gradient of the loss with respect to that party's embedding. Party~A receives $g_A = \partial L / \partial z_A$ and backpropagates it through $f_A$. The raw features never leave a party, and the labels never leave the server. Figure~\ref{fig:system} shows the setup, the attacker, and where the detector and the response sit on the server.

Following Fu et al.~\cite{fu2022label}, we split image inputs by pixel columns. With the split point at column~16, Party~A holds columns 0 to 15 and Party~B holds columns 16 to 31. Both bottom models are ResNet networks~\cite{he2016resnet}. The label party trains for 100 epochs on the ten-class datasets and 150 epochs on the hundred-class dataset.

\subsection{The active label inference attack}
The attack has two stages~\cite{fu2022label}. In the first stage, Party~A takes part in normal VFL training but replaces the optimizer of its bottom model with MaliciousSGD. For each parameter, MaliciousSGD compares the current parameter gradient to the previous one and scales it up when the two agree:
\begin{equation}
r = \mathrm{clamp}\!\left(1 + \gamma \frac{g^{(t)}}{g^{(t-1)}},\; 1,\; 5\right),\qquad p.\mathrm{grad} \leftarrow r \cdot p.\mathrm{grad},
\end{equation}
with $\gamma = 1$ and the ratio capped at 5 in our experiments. The scaling acts on the internal parameter gradient $p.\mathrm{grad}$, not on the communicated gradient $g_A$. Over many epochs this drives $f_A$ to produce embeddings whose classes form tight, well-separated clusters, even though Party~A never sees a label. A benign Party~A uses plain SGD for the same model, so the choice of optimizer is the only difference between the honest and adversarial runs.

\begin{figure}[t]
\centering
\begin{tikzpicture}[font=\small,
  box/.style={draw,rounded corners,align=center,inner sep=3pt,text width=1.95cm,minimum height=1.15cm},
  stg/.style={box,fill=gray!10},
  outbox/.style={box,fill=white},
  ar/.style={-{Stealth[length=2mm]},thick}]
  \node[stg] (s1) {\textbf{Stage 1}\\ joint VFL training; Party~A runs MaliciousSGD};
  \node[stg,right=10mm of s1] (s2) {\textbf{Stage 2}\\ freeze $f_A$, add head, MixMatch};
  \node[outbox,right=4mm of s2] (lab) {recovered labels};
  \draw[ar] (s1) -- node[above,font=\footnotesize]{save $f_A$} (s2);
  \draw[ar] (s2) -- (lab);
  \node[below=6mm of s2,font=\footnotesize,align=center] (aux) {$n_{\text{lab}}$ labels\\ $+$ unlabeled set};
  \draw[ar] (aux) -- (s2);
\end{tikzpicture}
\caption{The two-stage active attack. Stage~1 is normal VFL training, except Party~A runs MaliciousSGD, which shapes its bottom model $f_A$ into a strong class separator. Stage~2 works offline: the attacker freezes $f_A$, attaches a classifier head, and recovers the training labels with MixMatch from a small labeled set and the unlabeled data.}
\label{fig:pipeline}
\end{figure}
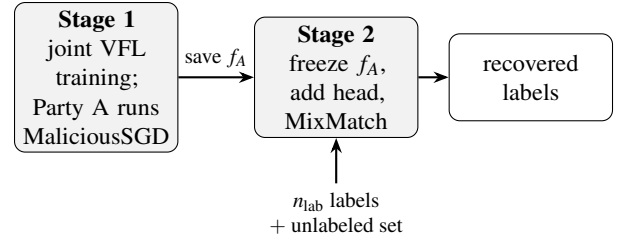

In the second stage, called model completion, Party~A works offline. It freezes the bottom model $f_A$ it saved from the first stage, attaches a small classifier head, and trains the head with the MixMatch semi-supervised method~\cite{berthelot2019mixmatch}. This uses a small labeled set that the attacker is assumed to hold, 40 examples for the ten-class datasets and 400 for the hundred-class dataset, together with the unlabeled training set. Because $f_A$ already separates the classes, a few labels are enough to name each cluster, and the head recovers the labels of the full training set. We report this training-set recovery rate as the leakage, since the attacker's goal is to label the users it shares with the label party. Figure~\ref{fig:pipeline} shows the two stages end to end.

\subsection{Threat model}
\noindent\textbf{Attacker.} One feature-holding party, Party~A. It follows the VFL protocol at the message level: it sends valid embeddings and consumes the gradients it receives. Its only deviations are local and hidden from the server. It swaps in MaliciousSGD during training, and it runs model completion offline after training. It does not control the server, does not observe Party~B's features, and does not see the labels. It holds a small labeled subset of its own samples, which is a mild assumption for a real participant.

\noindent\textbf{Threat surface.} The attacker's local optimizer and its saved bottom model. The attacker never has to break the protocol, so the attack is invisible to any check on message format or protocol order. The only externally visible effect is statistical: the embeddings Party~A sends have unusual class structure. The server observes these embeddings each batch and controls the gradients it returns, which is the surface our defense uses.

\noindent\textbf{Defender.} The server, which holds the labels and wants to keep them private while preserving task accuracy. The server sees each party's per-batch embeddings and the gradients it returns. It does not see any party's weights, optimizer state, or raw features. This matches a real deployment in which the label party runs the top model.

\noindent\textbf{Generality and practicality.} The attack needs only a local optimizer change and a few dozen labels, so it is cheap and does not depend on the modality of the data. The defense must be just as cheap and must not require the honest parties to change anything. Ours runs entirely on the server and adds one embedding statistic per batch.

\noindent\textbf{Scope.} We target the single adversarial party case, where the asymmetry between parties is the signal. If every party runs MaliciousSGD, the asymmetry disappears and our detector does not fire. We treat that case as out of scope and return to it in Section~\ref{sec:discussion}.

\subsection{Why existing defenses fall short}
\label{sec:existing}
The standard protections against gradient leakage perturb the returned gradient. Differential privacy adds calibrated noise~\cite{dwork2006calibrating}, gradient compression keeps only the largest entries~\cite{lin2018dgc}, and quantization rounds them. All of these act on the communicated gradient $g_A$ and apply the same perturbation to every party. Two problems follow. First, they are symmetric, so they degrade the honest party's learning as much as the attacker's, which costs task accuracy. Second, MaliciousSGD amplifies the internal gradient $p.\mathrm{grad}$, which is derived from $g_A$ through the attacker's own network. Weakening the message slows the attacker but does not stop the internal amplification from rebuilding class structure over a long training run. Our own experiments confirm this: noise and compression on $g_A$ leave a large share of the leakage in place. These limits motivate a defense that first detects which party is attacking and then acts only on that party.

\section{Attack Characterization: The Fisher Divergence Signal}
\label{sec:signal}

Our defense rests on one observation: the active attack changes the shape of the attacker's embedding space in a way the server can measure. This section defines the measurement, shows how it separates attack from honest training, and explains why a more obvious signal does not work.

\subsection{Per-party Fisher separability}
The server holds the labels, so it can measure how well each party's embeddings separate the classes. We use the Fisher ratio from linear discriminant analysis~\cite{fisher1936lda}. For a set of embeddings with class means $\mu_c$ and global mean $\mu$, let
\begin{equation}
J = \frac{\mathrm{Tr}(S_B)}{\mathrm{Tr}(S_W)},\quad
S_B = \sum_c n_c (\mu_c - \mu)(\mu_c - \mu)^\top,\
S_W = \sum_c \sum_{i \in c} (z_i - \mu_c)(z_i - \mu_c)^\top,
\end{equation}
where $S_B$ is the between-class scatter and $S_W$ is the within-class scatter. A high $J$ means the classes sit in tight clusters that are far apart. The server accumulates the embeddings it receives over an epoch and computes $J_A$ for Party~A and $J_B$ for Party~B. It needs nothing from the parties beyond the embeddings they already send.

We track the difference between the two parties,
\begin{equation}
\Delta_F = J_A - J_B.
\end{equation}
Party~B is honest by assumption, so $J_B$ acts as a live reference for what normal separability looks like on this task at this point in training. This is what makes $\Delta_F$ usable as a fixed-threshold signal across datasets: it measures the attacker against a moving baseline rather than against an absolute number.

\subsection{Why the attack raises the signal}
MaliciousSGD amplifies Party~A's parameter gradients whenever consecutive updates agree, which speeds up the formation of class structure in $f_A$. Party~B learns at the normal rate. So under the attack $J_A$ climbs faster and further than $J_B$, and $\Delta_F$ becomes and stays positive. In honest training both parties learn at similar rates and $\Delta_F$ stays near zero.

\begin{figure}[t]
\centering
\includegraphics[width=\linewidth]{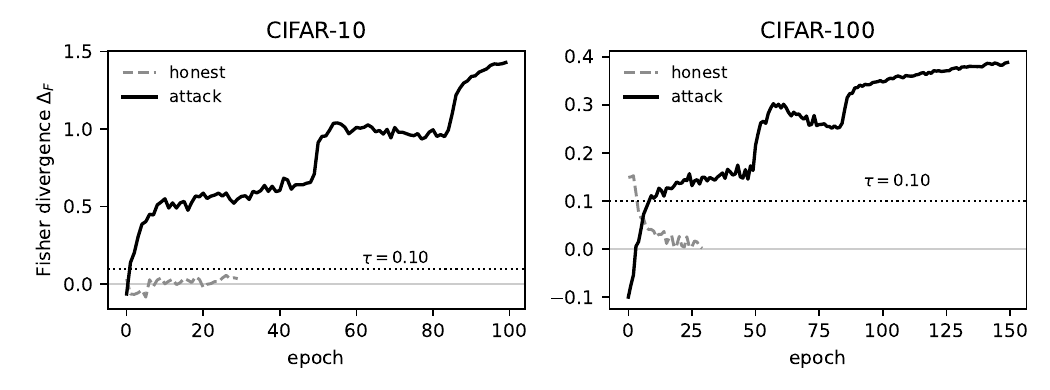}
\caption{Fisher divergence $\Delta_F = J_A - J_B$ over training. Honest training (shown over its 30-epoch characterization run) stays near zero. The active attack rises and stays positive, crossing the detection threshold (dotted, $\tau=0.10$) within a few epochs on both datasets.}
\label{fig:fisher}
\end{figure}

Figure~\ref{fig:fisher} shows the trajectories. On CIFAR-10, honest training holds $\Delta_F$ between $-0.14$ and $-0.10$ across the run, while the attack pushes it to $+0.44$ by epoch~29 and above $+1.2$ by epoch~99. On CIFAR-100 the gap is smaller but still clear: honest training stays near zero, and the attack settles between $+0.08$ and $+0.13$ after the early epochs. The attack signal is available early. It crosses a threshold of $0.10$ by the fourth epoch on CIFAR-10 and by the seventh on CIFAR-100, long before the attacker has finished building exploitable structure.

\subsection{Threshold, burn-in, and detection quality}
\begin{figure}[t]
\centering
\includegraphics[width=\linewidth]{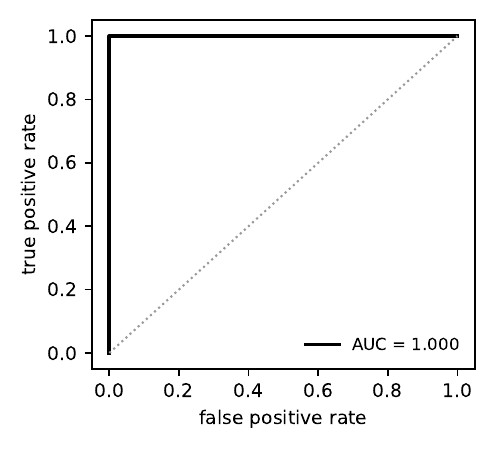}
\caption{Detection ROC for the Fisher divergence signal, per-epoch decisions on CIFAR-10 and CIFAR-100 after burn-in. Honest epochs are negatives and attack epochs are positives. The ranges are disjoint on these non-adaptive runs, so the area under the curve is 1.0.}
\label{fig:roc}
\end{figure}

Early in training the embeddings are unstable and $\Delta_F$ is noisy, so we hold the detector off for a short burn-in and then compare $\Delta_F$ against a fixed threshold $\tau$. We use $\tau = 0.10$ throughout. Table~\ref{tab:detection} gives the operating ranges. On the ten-class datasets the honest and attack ranges are far apart, so the threshold has a wide margin. On CIFAR-100 the two ranges are closer, which is the first sign that the hundred-class setting is harder, a point the defense design in Section~\ref{sec:defense} has to address. Because the honest and attack ranges do not overlap after burn-in, a fixed-threshold detector separates honest epochs from attacked epochs on these runs, and the per-epoch ROC over both datasets has an area under the curve of 1.0 (Figure~\ref{fig:roc}). This is the separation against a non-adaptive attacker; Section~\ref{sec:discussion} discusses an attacker that tries to hold its divergence below the threshold.

\begin{table}[t]
\centering
\small
\begin{tabular}{lccc}
\toprule
Dataset & Honest $\Delta_F$ & Attack $\Delta_F$ & Burn-in \\
\midrule
CIFAR-10  & $-0.14$ to $-0.10$ & $+0.33$ to $+0.53$ & 4 \\
CIFAR-100 & $+0.00$ to $+0.04$ & $+0.08$ to $+0.13$ & 8 \\
\bottomrule
\end{tabular}
\caption{Operating ranges of the detection signal after burn-in (epochs 8 to 29). The threshold $\tau = 0.10$ separates the two ranges on both datasets, with a wider margin for ten classes.}
\label{tab:detection}
\end{table}

\subsection{Why not the gradient norm}
A simpler idea is to watch the norm of the returned gradient and flag a party whose gradient looks unusual. This does not work here. MaliciousSGD amplifies the internal parameter gradient $p.\mathrm{grad}$, not the communicated gradient $g_A$ that the server sees. In our runs the ratio of the two parties' communicated gradient norms stays close to one under attack, so a norm-based detector has almost nothing to key on. The Fisher signal works because it measures the effect of the attack on the embedding geometry, which is exactly what the internal amplification changes, rather than the channel the amplification does not touch.

\subsection{Robustness and a known blind spot}
The signal survives a gradient-space defense running at the same time. When Party~A also applies gradient compression, $\Delta_F$ is if anything larger, because compression removes noise from the embedding without removing the class structure. The one case the detector misses by design is the all-parties attack, where Party~B also runs MaliciousSGD. Then both separabilities rise together, $\Delta_F$ stays near zero, and the detector does not fire. This is the cost of using cross-party asymmetry as the signal, and we discuss it in Section~\ref{sec:discussion}.

\section{Defense Design}
\label{sec:defense}

The detector tells the server when an attack is running and which party is running it. The defense uses both facts. It acts only while $\Delta_F$ is above the threshold, and it acts only on the attacker's gradient. The honest party is never touched, which is why the cost to the shared task stays small. What the server does to the attacker's gradient depends on the number of classes, and Section~\ref{sec:dcr} explains why the two cases differ. Algorithm~\ref{alg:defense} gives the full server-side procedure, from the per-batch gradient response to the per-epoch divergence update.

\subsection{Detection gate and asymmetric response}
On each batch, after the burn-in, the server checks whether the divergence from the previous epoch is above $\tau$. If it is, the server modifies $g_A$ before returning it to Party~A and leaves $g_B$ unchanged. If $\Delta_F$ is at or below $\tau$, the server returns both gradients unchanged. Because the modification is gated on $\Delta_F$, it stays off during honest training, where the divergence never crosses the threshold. This is what keeps the false-positive cost near zero.

\subsection{Adaptive suppression for few-class tasks}
The first response scales the attacker's gradient down in proportion to how far the divergence sits above the threshold:
\begin{equation}
s = \max\!\big(0,\; 1 - \alpha (\Delta_F - \tau)\big),\qquad g_A \leftarrow s\, g_A .
\end{equation}
At $\Delta_F = \tau$ the scale is 1 and nothing changes. As the divergence grows the scale falls, and it reaches 0 once $\Delta_F \ge \tau + 1/\alpha$. With $\alpha = 1$ and $\tau = 0.10$ the gradient is fully blocked once the divergence passes $1.1$.

This works on the ten-class datasets because the attack drives $\Delta_F$ well past $1.1$ in the second half of training. For those epochs Party~A receives no gradient at all, so its bottom model stops receiving the task signal it needs to align its clusters with the true labels. The shared task does not collapse, because Party~B's gradient is untouched and the top model keeps learning from it. On CIFAR-10 the attacker's gradient is fully blocked from about epoch~74 onward, yet the task loses under two points of test accuracy.

\subsection{Bounded projection for many-class tasks}
Suppression fails on a hundred classes. There the divergence peaks near $0.45$ and never reaches $\tau + 1/\alpha$, so the scale never reaches 0 and the attacker always gets some signal. Raising $\alpha$ does not help and makes things worse: cutting the task gradient early lets the attacker's internal amplification run without the corrective task signal, which produces sharper class clusters, not weaker ones. We confirm this in Section~\ref{sec:failed}.

Instead of shrinking the whole gradient, the server removes the direction that carries the class information. It trains a small auxiliary linear classifier $h$ on Party~A's embeddings, using the labels it already holds. The gradient of the auxiliary loss with respect to the embedding,
\begin{equation}
d = \frac{\partial L_{\mathrm{aux}}}{\partial z_A},
\end{equation}
points along the direction that makes $z_A$ more class-discriminative. The server removes the component of $g_A$ along $\hat d = d / \lVert d\rVert$:
\begin{equation}
g_{\mathrm{proj}} = g_A - (g_A^\top \hat d)\, \hat d .
\end{equation}
The auxiliary classifier is trained every batch so that $\hat d$ tracks the current discriminative direction, and the projection is applied only while the detector is firing.

\subsection{The projection cannot amplify}
The projection removes a component and never adds one, so
\begin{equation}
\lVert g_{\mathrm{proj}}\rVert = \lVert g_A\rVert \sin\theta \le \lVert g_A\rVert ,
\end{equation}
where $\theta$ is the angle between $g_A$ and $\hat d$. This bound is the reason the projection is safe under MaliciousSGD. Even when the attacker amplifies the projected gradient by the maximum factor of 5, the result is no larger than five times the original gradient, the same ceiling as an undefended run. An earlier design of ours removed the bound. It subtracted a scaled copy of $d$ from $g_A$ rather than projecting, which adds force instead of only removing it. Under amplification the auxiliary gradient grew without limit, the logits overflowed, and training diverged to not-a-number on every setting we tried. The projection form has no such failure mode, and we report the unbounded version among the failed defenses in Section~\ref{sec:failed}.

\subsection{Why the two cases differ: the Discriminative Coverage Ratio}
\label{sec:dcr}
Removing one direction removes a fixed slice of a subspace whose size grows with the number of classes. Linear discriminant analysis puts the class-separating information in a subspace of dimension $\min(C-1, m)$, where $C$ is the number of classes and $m$ is the embedding dimension. We define the Discriminative Coverage Ratio of a rank-$K$ projection as
\begin{equation}
\mathrm{DCR}(K, C) = \frac{K}{\min(C-1,\, m)} .
\end{equation}
For ten classes, a single direction covers $1/9 \approx 11\%$ of the discriminative subspace. For a hundred classes it covers $1/99 \approx 1\%$. This is the quantity that separates the two datasets. On ten classes, removing one direction, or suppressing the whole gradient once the divergence is large, takes out enough of the discriminative subspace to keep the clusters from aligning with the labels. On a hundred classes, one direction leaves 99 others open, and a sustained rank-one projection covers too little to hold the attacker back on its own.

On CIFAR-100 a single well-aimed projection is enough to drive leakage below the honest baseline, through a decisive one-step removal of the class-aligned gradient that we describe in Section~\ref{sec:mechanism}. DCR characterizes how projection coverage grows with the rank, and it points to a higher-rank projection, $K \approx (C-1)/9$ to match the ten-class ratio, as the route to even larger label spaces. We return to this extension in Section~\ref{sec:discussion}.

\begin{algorithm}[t]
\caption{VFL training with the detection-triggered asymmetric defense (server side).}
\label{alg:defense}
\begin{algorithmic}[1]
\Require bottom models $f_A, f_B$; top model $\mathrm{Top}$; labels $y$; class count $C$; threshold $\tau$; strength $\alpha$; burn-in $T_b$; epochs $E$
\State initialize auxiliary classifier $h$; \quad $\Delta_F \gets 0$
\For{epoch $t = 1$ to $E$}
  \State clear epoch buffers $Z_A, Z_B, Y$
  \For{each batch $(x_A, x_B, y)$}
    \State $z_A \gets f_A(x_A)$; \quad $z_B \gets f_B(x_B)$ \Comment{parties send embeddings}
    \State $L \gets \mathrm{CE}(\mathrm{Top}(z_A, z_B),\, y)$
    \State $g_A \gets \partial L / \partial z_A$; \quad $g_B \gets \partial L / \partial z_B$
    \State train $h$ on $(z_A, y)$; \; $d \gets \partial\, \mathrm{CE}(h(z_A), y)/\partial z_A$; \; $\hat d \gets d/\lVert d\rVert$
    \If{$t \ge T_b$ \textbf{and} $\Delta_F > \tau$} \Comment{detection gate}
      \If{$C$ is small} \Comment{adaptive suppression}
        \State $g_A \gets \max\!\big(0,\, 1 - \alpha(\Delta_F - \tau)\big)\, g_A$
      \Else \Comment{bounded projection}
        \State $g_A \gets g_A - (g_A^\top \hat d)\, \hat d$
      \EndIf
    \EndIf
    \State send $g_A$ to A, $g_B$ to B \Comment{honest party untouched}
    \State parties update $f_A, f_B$ \Comment{A uses MaliciousSGD}
    \State append $z_A, z_B, y$ to $Z_A, Z_B, Y$
  \EndFor
  \State $J_A \gets \mathrm{Fisher}(Z_A, Y)$; \quad $J_B \gets \mathrm{Fisher}(Z_B, Y)$
  \State $\Delta_F \gets J_A - J_B$ \Comment{used in the next epoch}
\EndFor
\end{algorithmic}
\end{algorithm}

\subsection{Analysis: boundedness, coverage, and cost}

\noindent\textbf{The projection cannot amplify.} Let $g$ be any gradient and $\hat d$ a unit vector. The projection $Pg = g - (g^\top \hat d)\,\hat d$ satisfies
\begin{equation}
\lVert Pg\rVert^2 = \lVert g\rVert^2 - (g^\top \hat d)^2 \le \lVert g\rVert^2 ,
\end{equation}
with equality only when $g \perp \hat d$, so $\lVert Pg\rVert \le \lVert g\rVert$ for every $g$. When Party~A amplifies the projected gradient by its factor $r \le 5$, the result is bounded by $5\lVert g\rVert$, the same ceiling as an undefended run. This is why the projection never diverges, unlike the unbounded reversal in Section~\ref{sec:failed}, which adds a term to the gradient and can grow without limit.

\noindent\textbf{Coverage scales with the rank.} The class-separating information lies in a subspace of dimension at most $\min(C-1, m)$, where $m$ is the embedding dimension. A rank-$K$ projection removes $K$ of its directions, so it covers $\mathrm{DCR}(K,C) = K / \min(C-1, m)$ of that subspace. To hold a fixed coverage $\kappa$ as the class count grows, the rank must grow with it, $K = \lceil \kappa \cdot \min(C-1, m) \rceil$. The defense we evaluate is the $K = 1$ case, which is enough for ten classes ($\mathrm{DCR} \approx 11\%$) but thin for a hundred ($\mathrm{DCR} \approx 1\%$).

\noindent\textbf{Overhead.} Detection adds one Fisher ratio per party per epoch. From $N$ embeddings of dimension $m$ over $C$ classes, the class means and scatter traces cost $O(Nm + Cm)$, and the server can stream these statistics rather than store every embedding. The projection response adds one linear auxiliary classifier of size $m \times C$, trained one step per batch at $O(mC)$. Neither party changes anything, and the server never touches any party's weights or optimizer state.

\section{Evaluation}
\label{sec:eval}

We evaluate four questions. Does the defense reduce label leakage below what honest training leaks? What does it cost the shared task? How does it actually work? And which simpler defenses fail, and why?

\subsection{Setup}
We use three image datasets: CIFAR-10 and CIFAR-100~\cite{krizhevsky2009cifar}, and CINIC-10~\cite{darlow2018cinic}. CIFAR-10 and CINIC-10 have ten classes; CIFAR-100 has a hundred. Each input is split by pixel column at position 16, so Party~A and Party~B each hold half. Both bottom models are ResNet networks~\cite{he2016resnet}. The label party trains for 100 epochs on the ten-class datasets and 150 on CIFAR-100. The attacker's model completion runs for 25 epochs of MixMatch~\cite{berthelot2019mixmatch} with 40 labeled examples on the ten-class datasets and 400 on CIFAR-100. We report the best training-set top-1 recovery over the model completion run as the leakage, since the attacker's goal is to label the shared users. Every condition is run over four seeds ($0, 42, 123, 456$), and we report the mean and standard deviation. The defense uses $\alpha = 1$, $\tau = 0.10$, and a burn-in of 8 on every dataset, with no per-dataset tuning. Table~\ref{tab:setup} summarizes the datasets and settings.

\begin{table}[t]
\centering
\small
\begin{tabular}{lcccc}
\toprule
Dataset & Classes & Train & $n_{\text{lab}}$ & Stage-1 ep. \\
\midrule
CIFAR-10  & 10  & 50{,}000 & 40  & 100 \\
CINIC-10  & 10  & 90{,}000 & 40  & 100 \\
CIFAR-100 & 100 & 50{,}000 & 400 & 150 \\
\bottomrule
\end{tabular}
\caption{Datasets and settings. Inputs are split by pixel column at position 16, so each party holds half. $n_{\text{lab}}$ is the number of labeled examples available to model completion, which runs for 25 MixMatch epochs.}
\label{tab:setup}
\end{table}

\subsection{Main results}
Table~\ref{tab:main} gives the headline numbers. On all three datasets the defended leakage sits below the leakage of honest training, and the shared task loses little or no accuracy. The attack raises CIFAR-10 leakage from the honest 83.11\% to 94.95\%; the defense brings it to 81.80\%, which is below honest. On CINIC-10 the attack advantage is larger, 86.51\% against an honest 65.76\%, and the defense still lands at 62.72\%, below honest. On CIFAR-100 the attack reaches 49.87\% against an honest 29.56\%, and the projection defense brings it to 23.44\%. The ten-class datasets use adaptive suppression and the hundred-class dataset uses the bounded projection, following Section~\ref{sec:defense}. The cost to the joint task is under two points of test accuracy in every case, and on CIFAR-100 and CINIC-10 it is within noise.

\begin{table*}[t]
\centering
\small
\begin{tabular}{llccccc}
\toprule
Dataset & Classes & Honest & Attack & Defended & Defended $-$ Honest & Task cost \\
\midrule
CIFAR-10  & 10  & $83.11 \pm 2.84$ & $94.95 \pm 0.52$ & $81.80 \pm 1.85$ & $-1.31$ & $-1.9$ \\
CINIC-10  & 10  & $65.76 \pm 0.65$ & $86.51 \pm 0.54$ & $62.72 \pm 0.65$ & $-3.04$ & $+0.6$ \\
CIFAR-100 & 100 & $29.56 \pm 2.93$ & $49.87 \pm 1.17$ & $23.44 \pm 5.16$ & $-6.12$ & $\approx 0$ \\
\bottomrule
\end{tabular}
\caption{Label inference (best training-set top-1, \%) over four seeds. Defended leakage is below honest training on all three datasets. Ten-class datasets use adaptive suppression; CIFAR-100 uses the bounded projection. Task cost is the change in VFL test accuracy against honest training.}
\label{tab:main}
\end{table*}

Within the CIFAR-100 defended column, three of the four seeds cluster between 25\% and 27\%, and the fourth is suppressed further, to 14.6\%. All four sit below the honest baseline. The fourth seed carries the largest task-accuracy cost of the group, and we include it in the reported mean rather than dropping it.

\subsection{Reproducing the attack}
Our attack reproduces the strong active-attack leakage reported by Fu et al.~\cite{fu2022label}. Table~\ref{tab:attackrepro} gives the advantage of the active attack over the passive baseline in our runs: the active attacker recovers 12 to 21 more points of the training labels than a passive party on the same setup. This confirms that our defense is tested against a working attack, not a weakened one. The corresponding figures from the original paper can be read into the last column for a direct comparison.

\begin{table}[t]
\centering
\small
\begin{tabular}{lcccc}
\toprule
Dataset & Passive & Active & Advantage & Fu et al.~\cite{fu2022label} \\
\midrule
CIFAR-10  & 83.11 & 94.95 & $+11.84$ & $\dagger$ \\
CINIC-10  & 65.76 & 86.51 & $+20.75$ & -- \\
CIFAR-100 & 29.56 & 49.87 & $+20.31$ & $\dagger$ \\
\bottomrule
\end{tabular}
\caption{Active label inference over the passive baseline, our measurements (best training-set top-1, \%, four-seed means). The active attack has a large advantage on every dataset. $\dagger$: insert the value reported in~\cite{fu2022label}; CINIC-10 was not studied there.}
\label{tab:attackrepro}
\end{table}

\subsection{Per-seed results}
Table~\ref{tab:perseed} gives the per-seed leakage behind the four-seed means in Table~\ref{tab:main}. B is the passive (honest) baseline, A is the active attack, and D is the defended run. The defended column is below the passive column for every seed on every dataset. CIFAR-100 seed 456 is suppressed the furthest, to 14.57\%, which widens that column's standard deviation while staying well below the honest baseline.

\begin{table*}[t]
\centering
\small
\begin{tabular}{lccccccccc}
\toprule
& \multicolumn{3}{c}{CIFAR-10} & \multicolumn{3}{c}{CINIC-10} & \multicolumn{3}{c}{CIFAR-100} \\
\cmidrule(lr){2-4}\cmidrule(lr){5-7}\cmidrule(lr){8-10}
Seed & B & A & D & B & A & D & B & A & D \\
\midrule
0   & 87.23 & 95.42 & 84.27 & 65.70 & 86.59 & 62.43 & 30.33 & 47.86 & 26.97 \\
42  & 81.90 & 94.32 & 80.16 & 66.12 & 85.94 & 63.08 & 26.19 & 50.40 & 25.33 \\
123 & 80.78 & 95.33 & 80.61 & 64.88 & 87.21 & 61.95 & 28.56 & 50.73 & 26.89 \\
456 & 82.53 & 94.73 & 82.14 & 66.35 & 86.31 & 63.41 & 33.14 & 50.50 & 14.57 \\
\midrule
Mean & 83.11 & 94.95 & 81.80 & 65.76 & 86.51 & 62.72 & 29.56 & 49.87 & 23.44 \\
Std  & 2.84 & 0.52 & 1.85 & 0.65 & 0.54 & 0.65 & 2.93 & 1.17 & 5.16 \\
\bottomrule
\end{tabular}
\caption{Per-seed label inference (best training-set top-1, \%). B: passive baseline. A: active attack. D: defended. The defended value is below the passive value for every seed on every dataset.}
\label{tab:perseed}
\end{table*}

\subsection{The defense stays off during honest training}
A defense that fires on honest training would tax every deployment, not just attacked ones. Ours does not fire, because the detection gate never opens. On CIFAR-10 the honest divergence stays below $\tau$ for the whole run, so the scale stays at 1 and no gradient is changed. Running the full pipeline on honest CIFAR-10 with the defense active gives 84.35\% leakage, within the honest band of $83.11 \pm 2.84\%$. On CIFAR-100 the honest divergence goes negative after burn-in, so the projection never activates. The defense is a no-op on honest training by construction, which is why the task cost in Table~\ref{tab:main} comes only from attacked runs.

\subsection{How the defense works}
\label{sec:mechanism}
The defense does not protect the labels by destroying the embedding geometry. It protects them by breaking the link between the geometry and the labels.

\begin{figure}[t]
\centering
\includegraphics[width=\linewidth]{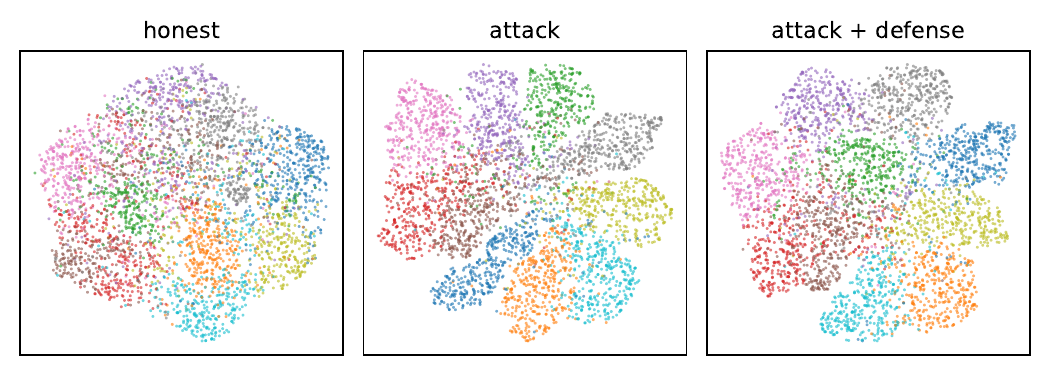}
\caption{t-SNE of Party~A's embeddings on CINIC-10, colored by true class. The attack produces tight, well-separated clusters. Under the defense the embedding keeps visible class structure rather than collapsing into noise, yet label recovery still falls below honest training (Table~\ref{tab:main}), which we read as a loss of cluster-label alignment.}
\label{fig:tsne}
\end{figure}

On CIFAR-10 under suppression, the attacker's Fisher separability stays high. At the last epoch $J_A$ is 1.74, more than three times the honest value, so the clusters are still there and still tight. Yet the training-set leakage is 81.80\%, just below the honest 83.11\%. The defense does not remove the geometry. It weakens how well that geometry tracks the true labels. MixMatch recovers labels by spreading a few seed labels across the embedding, so even a partial loss of alignment between the clusters and the labels lowers the recovery rate. Figure~\ref{fig:tsne} is consistent with this reading. Under the defense the embedding keeps visible class structure rather than collapsing into noise, which rules out the simpler explanation that the defense works by destroying the embedding. We read the drop below honest training as a loss of cluster-label alignment rather than a loss of structure, and we note that the CIFAR-10 margin here is small; the effect is clearer on CINIC-10, where the defended leakage is three points below honest.

\begin{figure}[t]
\centering
\includegraphics[width=\linewidth]{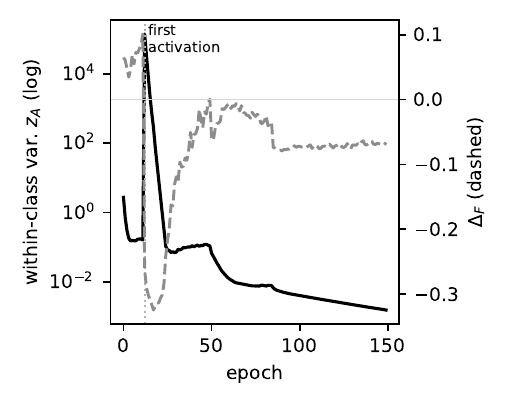}
\caption{CIFAR-100 projection defense. At the first activation the within-class variance of Party~A rises sharply and the Fisher divergence turns negative. Party~A stays the less discriminative party for the rest of training, so the detector does not fire again.}
\label{fig:collapse}
\end{figure}

On CIFAR-100 the projection acts more decisively. The auxiliary classifier spends the burn-in learning an accurate discriminative direction. When the detector first fires, that direction is almost parallel to $g_A$, so the projection removes nearly all of the class-aligned gradient in one step. Party~A's bottom model does not rebuild its class structure afterward: its Fisher separability drops below Party~B's and stays there for the rest of training, the detector does not need to fire again, and the saved bottom model no longer separates the classes. Figure~\ref{fig:collapse} shows this. The effect is consistent across seeds and costs almost nothing in task accuracy, because one well-aimed projection removes the attacker's advantage without repeated intervention. The DCR analysis in Section~\ref{sec:dcr} places this in a general picture: projection coverage grows with the rank, and a rank that scales with the class count carries the same idea to even larger label spaces.

\subsection{Sensitivity to the defense parameters}
Table~\ref{tab:ablation} sweeps the suppression strength $\alpha$ and the threshold $\tau$ on CIFAR-10. Four of the five settings bring leakage below the honest baseline. The one failure is the deliberately weak $\alpha = 0.5$, which barely suppresses the gradient. Stronger suppression is safe here: $\alpha = 2$ pushes leakage more than ten points below honest. Lowering the threshold to $0.05$ or raising it to $0.15$ both stay below honest, so the defense does not depend on a fragile threshold choice for ten-class tasks.

\begin{table}[t]
\centering
\small
\begin{tabular}{lccc}
\toprule
Setting & Leakage & vs Honest & Task cost \\
\midrule
$\alpha=0.5,\ \tau=0.10$ & 94.57 & $+7.3$ & $-0.4$ \\
$\alpha=1.0,\ \tau=0.10$ & 84.27 & $-3.0$ & $-1.9$ \\
$\alpha=2.0,\ \tau=0.10$ & 76.79 & $-10.4$ & $-1.9$ \\
$\alpha=1.0,\ \tau=0.05$ & 81.58 & $-5.7$ & $-3.2$ \\
$\alpha=1.0,\ \tau=0.15$ & 85.98 & $-1.3$ & $-2.5$ \\
\bottomrule
\end{tabular}
\caption{Parameter sweep on CIFAR-10 (single seed, honest reference 87.23\%). Four of five settings hold leakage below honest; only the deliberately weak $\alpha=0.5$ fails.}
\label{tab:ablation}
\end{table}

\subsection{What did not work}
\label{sec:failed}
We reached the two-part design by ruling out simpler ideas. Table~\ref{tab:failed} lists the defenses that failed on CIFAR-100, the hardest case. Suppression alone leaves leakage 13 points above honest. Stronger suppression is worse, not better, because cutting the task gradient early lets the attacker's internal amplification build sharper clusters. Adding calibrated noise to the returned gradient, or to the embedding before the top model, both push leakage above the undefended attack, because the top model responds to a noisy embedding with a more class-informative correction that the attacker then amplifies. Flipping the sign of the returned gradient, meant to cancel the amplification, fails because the amplification acts on the internal gradient, whose sign does not follow the returned one through the network. The unbounded direction-reversal method, which subtracts a scaled copy of the discriminative direction instead of projecting, diverges to not-a-number and destroys the model, as noted in Section~\ref{sec:defense}. A persistent single-direction projection that re-applies every detected epoch does not reach below-honest leakage on either class count, which is consistent with the DCR view that a single direction covers only a small fraction of the discriminative subspace.

\begin{table}[t]
\centering
\small
\begin{tabular}{lcc}
\toprule
Defense on CIFAR-100 & Leakage & vs Honest \\
\midrule
Suppression $\alpha=1.0$        & 43.12 & $+13.6$ \\
Suppression $\alpha=2.0$        & 48.31 & $+18.8$ \\
Returned-gradient noise         & 43.1 to 49.6 & $+13.5$ to $+20.1$ \\
Embedding noise                 & 50.7 to 51.6 & $+21$ to $+22$ \\
Unbounded reversal              & \multicolumn{2}{c}{diverges (NaN)} \\
Single-direction persistent proj. & 48.8 to 50.2 & $+19$ to $+20$ \\
\midrule
Bounded projection (ours)       & \textbf{23.4} & $\mathbf{-6.1}$ \\
\bottomrule
\end{tabular}
\caption{Defenses that fail on CIFAR-100 (best leakage \%, honest mean 29.56\%). Only the bounded projection brings leakage below honest. Sign-flip is omitted here; it fails on CIFAR-10 at 86.4\% against an 83.1\% honest baseline.}
\label{tab:failed}
\end{table}

\subsection{Comparison with symmetric perturbation}
Symmetric gradient defenses can suppress the attack, but only at a utility cost that our asymmetric defense avoids. Laplace noise strong enough to cut CIFAR-100 leakage also collapses the joint task: at the noise scale we tested, the VFL test accuracy falls to 2.87\%, close to random for a hundred classes. Gradient compression at 75\% does not suppress the attack at all and, because it removes noise from the embedding, slightly raises separability. Neither can tell which party is attacking, so neither can act selectively. Our defense reaches below-honest leakage while holding task accuracy, because it acts only on the attacker and only while the attack is running. We note that our symmetric baselines were run on a shorter training schedule than the main results, so we read them as indicative of the utility-versus-protection trade-off rather than as a matched head-to-head, and we mark a matched-schedule comparison as future work.

\section{Discussion and Limitations}
\label{sec:discussion}

\noindent\textbf{An adaptive attacker.} Our detector watches one statistic, so an attacker who knows this may try to evade it. There are two routes, and both have a cost. The attacker can hold its separability down to keep $\Delta_F$ below the threshold, but separability is the very thing that makes the labels recoverable, so staying under the threshold weakens the attack in step with the evasion. Or the attacker can remove the asymmetry by having both feature parties amplify, which is the all-parties case. That needs the second feature party to collude, which the honest-party assumption rules out, and it changes the threat model rather than defeating the defense within it. We do not claim robustness against a fully adaptive attacker who optimizes directly against the detector, and we see this as the main open problem for detection-based defenses of this kind.

\noindent\textbf{The all-parties blind spot.} The detector uses the gap between the two parties, so it cannot see an attack in which both parties behave the same way. This is a design choice, not an oversight: the asymmetry is what lets the defense act on the attacker alone and leave the honest party untouched. A defense for the symmetric case would need a different signal, likely an absolute rather than a relative measure of separability, and would give up the low false-positive cost that the asymmetry buys.

\noindent\textbf{Scaling to more classes.} On CIFAR-100 the projection reaches below-honest leakage through a single decisive removal of the class-aligned gradient, with three of four seeds in a tight band and the fourth suppressed further. The DCR framework places this on a general footing: a rank-$K$ projection covers $K/\min(C-1,m)$ of the discriminative subspace, so for label spaces beyond a hundred classes a higher-rank projection, $K \approx (C-1)/9$, keeps the coverage fixed. This is a direct extension of the projection we evaluate, and we leave its evaluation on larger label spaces to future work.

\noindent\textbf{Modality.} We evaluate on image datasets. The attack and the signal do not depend on the modality in principle, since both are defined on embeddings, but we do not have results for text or tabular VFL and do not claim them.

\noindent\textbf{Deployment cost.} The defense runs on the server. Detection adds one Fisher ratio per party per epoch, computed from embeddings the server already receives. The projection adds a small linear auxiliary classifier trained on those embeddings. Neither needs any change from the honest parties, and neither needs access to party weights or optimizer state.

\section{Related Work}

Table~\ref{tab:defcompare} places our defense among prior approaches. It is the only one that combines an attack detector, an asymmetric response that spares the honest party, and a rank that scales with the class count.

\begin{table}[t]
\centering
\footnotesize
\setlength{\tabcolsep}{4pt}
\begin{tabular}{lccccc}
\toprule
Method & Asym. & Detect & Active & Multi-cls & Cost \\
\midrule
DP noise~\cite{dwork2006calibrating,abadi2016deep} & \xmark & \xmark & \xmark & --- & high \\
Grad. comp.~\cite{lin2018dgc}      & \xmark & \xmark & \xmark & --- & low \\
MARVELL~\cite{li2022labelleakage}  & \xmark & \xmark & \xmark & \xmark & med \\
MixPro~\cite{fedads2023mixpro}     & \xmark & \xmark & \xmark & --- & low \\
ProjPert~\cite{projpert2024}       & \xmark & \xmark & \xmark & --- & med \\
LADSG~\cite{ladsg2025}             & ---    & \cmark & \cmark & --- & --- \\
VMask~\cite{vmask2025}             & \xmark & \xmark & \cmark & \cmark & crypto \\
\textbf{Ours}                      & \cmark & \cmark & \cmark & \cmark & low \\
\bottomrule
\end{tabular}
\caption{Where our defense sits among prior work. Asym.: acts on the attacker alone. Detect: uses an attack detector. Active: targets the active attack. Multi-cls: handles many classes. Entries reflect each cited paper's stated scope and our reading; ``---'' means not applicable or not reported.}
\label{tab:defcompare}
\end{table}

\noindent\textbf{Inference attacks on collaborative learning.} Training a model across parties opens several ways to recover private information. Membership inference tells whether a record was in the training set~\cite{shokri2017membership,nasr2019comprehensive}. Model inversion reconstructs input features from model outputs~\cite{fredrikson2015model}. In collaborative and federated training, the gradients shared during learning leak information about the data that produced them: Melis et al.~\cite{melis2019exploiting} recover unintended properties from shared updates, and a line of gradient-inversion work reconstructs training inputs from a single gradient~\cite{zhu2019deepleakage,zhao2020idlg,geiping2020inverting}. Label inference is the part of this space that targets the labels rather than the features, which is the asset the VFL label party holds.

\noindent\textbf{Label inference in split and vertical FL.} Fu et al.~\cite{fu2022label} introduced the passive, active, and direct label inference attacks that we defend against, including the MaliciousSGD optimizer and the model completion procedure. Pasquini et al.~\cite{pasquini2021tiger} study inference attacks on split learning~\cite{vepakomma2018split,gupta2018distributed} more broadly. Li et al.~\cite{li2022labelleakage} analyze label leakage from gradients in two-party split learning and propose MARVELL, a noise design that equalizes the class gradient distributions; MARVELL targets the binary case, where the noise design is tractable, while our setting has up to a hundred classes. Sun et al.~\cite{sun2022forward} study label leakage from the forward embedding, and Zou et al.~\cite{zou2022batch} address batch-level label inference and replacement. Together these establish label inference as a central risk for vertical federated learning~\cite{yang2019federated,liu2024vfl}.

\noindent\textbf{Gradient perturbation and cryptographic defenses.} The standard privacy tools perturb or hide the shared gradient. Differential privacy adds calibrated noise~\cite{dwork2006calibrating,dwork2014algorithmic,abadi2016deep}, gradient sparsification and compression drop or shrink entries~\cite{shokri2015privacy,lin2018dgc}, and secure aggregation hides individual updates behind a sum~\cite{bonawitz2017practical}. Cryptographic frameworks such as SecureBoost~\cite{cheng2021secureboost} and homomorphic-encryption schemes for vertically partitioned data~\cite{hardy2017private} protect the computation itself, and VMask~\cite{vmask2025} protects against model completion with layer masking and secret sharing. These approaches are either symmetric, so they cost the honest party as much as the attacker, or they change the protocol, and none of them detects which party is attacking.

\noindent\textbf{Projection and detection defenses.} MixPro, within the FedAds benchmark~\cite{fedads2023mixpro}, applies a per-batch gradient projection in VFL, but with a generic direction, against passive leakage, and without attack detection. ProjPert~\cite{projpert2024} uses ``projection'' to describe a search over noise parameters rather than a geometric projection onto a subspace. LADSG~\cite{ladsg2025} detects attacks from gradient-norm anomalies and responds by substituting gradients. Our defense differs on three points: the projection direction comes from an auxiliary classifier that tracks the discriminative subspace, the response is gated by a detector that measures embedding separability rather than gradient magnitude, and we tie the projection rank to the class count through the DCR.

\noindent\textbf{Semi-supervised learning and separability tools.} The attack's second stage recovers labels with semi-supervised learning through MixMatch~\cite{berthelot2019mixmatch}; related methods such as FixMatch~\cite{sohn2020fixmatch} and contrastive pretraining~\cite{chen2020simclr} would serve the same role, which is why a defense that degrades the label alignment of the embedding is more durable than one aimed at a single completion method. Our detection signal is the Fisher linear discriminant ratio~\cite{fisher1936lda,bishop2006pattern} computed on per-party embeddings, a different object from the Fisher information matrix used to calibrate noise in some private learning work. We visualize the embedding geometry with t-SNE~\cite{maaten2008tsne}.

\section{Conclusion}

The active label inference attack works through the attacker's own optimizer, which is why perturbing the shared gradient does not stop it. We showed that the attack leaves a signal the server can read from the embeddings it already receives: the attacker's Fisher separability pulls away from the honest party's, and the gap is visible within a few epochs. A defense triggered by this signal can act on the attacker alone. Scaling the attacker's gradient down handles few-class tasks, and a norm non-expansive projection handles many-class tasks, with the Discriminative Coverage Ratio explaining why the two cases need different ranks. Across three datasets and four seeds, the defense holds label leakage below what honest training already leaks, at under two points of task cost, and it does so by breaking the alignment between embedding clusters and labels rather than by destroying the embedding. The main open problem is an attacker that adapts directly against the detector, which we leave to future work.

\appendix

\section*{Ethical Considerations}

This work is defensive. The attack we study, including the MaliciousSGD optimizer and the model completion procedure, was introduced and published in prior work~\cite{fu2022label}; we do not propose a new attack or make the existing one stronger. Our contribution is a way for the label-holding party to detect and blunt that attack.

All experiments use public benchmark datasets (CIFAR-10, CIFAR-100, CINIC-10) that contain no personal or sensitive information and no human subjects. We ran everything in a controlled setting with simulated parties, so no real user data was exposed and no live system was attacked. The characterization of the attack in Section~\ref{sec:signal} could in principle help an attacker understand when its behavior is visible, but the same analysis is what makes the defense possible, and the net effect of publishing it favors the defender, who controls the server and can deploy the detector. We see no additional harm from releasing the detection signal and the defense together.

\section*{Open Science}

We will release all code, run scripts, and result files needed to reproduce every table and figure in this paper. The artifact includes the VFL training framework, the MaliciousSGD attack, the model completion procedure, the detection monitor, the defense implementations, and the scripts that produced each experiment, together with the raw per-epoch outputs behind Tables~\ref{tab:main} through~\ref{tab:failed} and Figures~\ref{fig:fisher} through~\ref{fig:collapse}.

The artifact is available at an anonymous repository for review:
\begin{center}
\texttt{[ANONYMIZED ARTIFACT URL: anonymous.4open.science/r/VFL-experiments-02D8]}
\end{center}
The repository contains no author or institution identifiers. On acceptance we will publish it at a stable, non-anonymous location.

\bibliographystyle{plainurl}
\bibliography{references}

\end{document}